\documentclass[10pt,aps,pre,twocolumn,final,letterpaper,superscriptaddress,longbibliography]{revtex4-1}

\usepackage[utf8]{inputenc}
\usepackage{graphicx}
\usepackage{amsmath}
\usepackage{txfonts}
\usepackage[hidelinks]{hyperref}

\graphicspath{{./figures/}}

\begin{document}

\author{Laurent H\'{e}bert-Dufresne}
\affiliation{Vermont Complex Systems Center, University of Vermont, Burlington, VT 05405 USA}
\affiliation{Department of Computer Science, University of Vermont, Burlington, VT 05405 USA}
\affiliation{D\'epartement de physique, de g\'enie physique et d'optique,
Universit\'e Laval, Qu\'ebec (Qu\'ebec), Canada G1V 0A6}

\author{M\'arton P\'osfai}
\affiliation{Department of Network and Data Science, Central European University, 1100 Vienna, Austria}

\author{Antoine Allard}
\affiliation{D\'epartement de physique, de g\'enie physique et d'optique,
Universit\'e Laval, Qu\'ebec (Qu\'ebec), Canada G1V 0A6}
\affiliation{Centre interdisciplinaire en mod\'elisation math\'ematique, Universit\'e Laval, Qu\'ebec (Qu\'ebec), Canada G1V 0A6}
\affiliation{Vermont Complex Systems Center, University of Vermont, Burlington, VT 05405 USA}

\title{Modeling critical connectivity constraints in random and empirical networks}

\begin{abstract}
Random networks are a powerful tool in the analytical modeling of complex networks as they allow us to write approximate mathematical models for diverse properties and behaviors of networks.  One notable shortcoming of these models is that they are often used to study processes in terms of how they affect the giant connected component of the network, yet they fail to properly account for that component.  As an example, this approach is often used to answer questions such as how robust is the network to random damage but fails to capture the structure of the network before any inflicted damage.  Here, we introduce a simple conceptual step to account for such connectivity constraints in existing models.  We distinguish network neighbors into two types of connections that can lead or not to a component of interest, which we call critical and subcritical degrees.  In doing so, we capture important structural features of the network in a system of only one or two equations.  In particular cases where the component of interest is surprising under classic random network models, such as sparse connected networks, a single equation can approximate state-of-the art models like message passing which require a number of equations linear in system size.  We discuss potential applications of this simple framework for the study of infrastructure networks where connectivity constraints are critical to the function of the system.
\end{abstract}

\maketitle

\section{Introduction: Models of networks}

Random networks are often used to model real networks while allowing simple mathematical analysis of their structure and properties.  Such models are typically parameterized with local connection rules enforcing structural constraints while leaving the rest of their structure random.  In theory, these rules should encode constraints that we believe capture important features of real networks while the randomness allow us to mathematically tract the model.  One of the most popular version of this approach is the \textit{configuration model} (CM), specified using the degree sequence of the network while assuming that edges are the connected at random~\cite{fosdick2018configuring}.  Many generalizations have been made, to include degree-degree correlations between neighbors~\cite{vazquez2003resilience} or more macroscopic structures like $k$-core or onion decompositions~\cite{hebert2013percolation, allard2019percolation}.

One notable example of this approach is the study of site or bond percolation on random networks, used to model the robustness of real networks to random failure or targeted attacks~\cite{albert2000error} and understand how they support dynamics like the spread of epidemics or cascading failures~\cite{newman2002spread}.  In these contexts, models can often approximate the percolation threshold of real network quite well, especially for denser networks~\cite{allard2019percolation}, but can fail to capture the actual size and robustness of the largest connected component well above the threshold.  Given that these models are often used to study the robustness of a structure to damage or to small changes in edge functionality, it is unfortunate that they often dramatically fail in the regime where there is no damage or changes to the structure.

Here, we aim to explore a new computationally-informed model by introducing the idea of critical connections.  In a nutshell, our goal is to specify a critical substructure of interest identified computationally---like the largest connected component---and model edges that contribute to this critical substructure separately from other edges of the network.  The resulting model can implicitly account for important network features, such as degree correlations or unknown statistics, that drive the connectivity of the substructure.  By moving some of the complexity of the problem to computational pre-processing, we hope to provide rich simple models to further the analytical study of complex networks.

\begin{figure}[b!]
    \centering
    \includegraphics[width=0.7\linewidth]{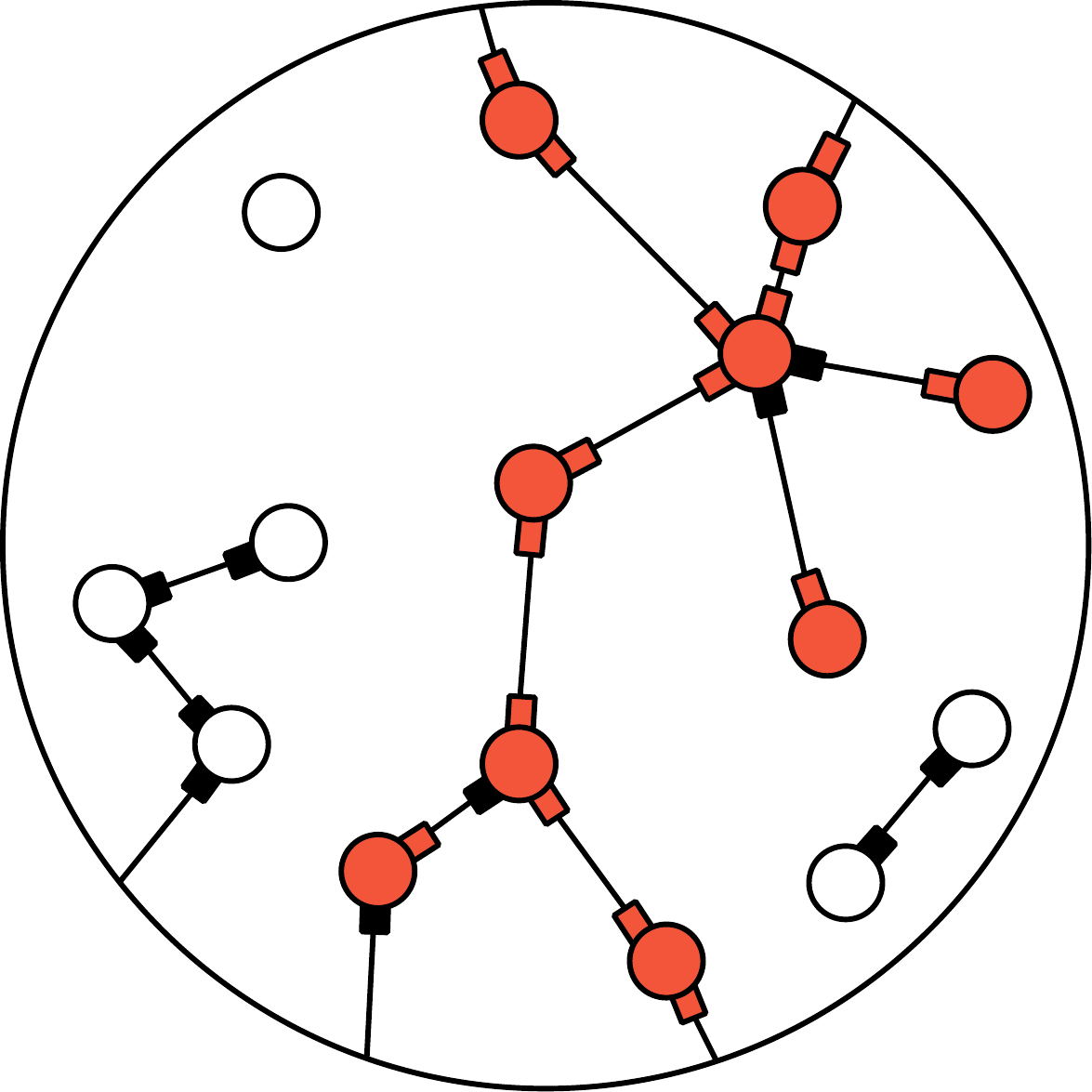}
    \caption{Illustration of a Critical Configuration Model informed by the giant component of a network.  Imagining all stubs as directed non-reciprocal edges, one can tag all stubs that point towards a certain fraction of the largest connected component (e.g., above 1\% of the system size) as contributing to the ``critical degree'' of nodes.  Nodes with a critical degree of at least 1 are then known to be in the giant component, but not all stubs in the giant component are critical as some lead to dead-ends.}
    \label{fig:cartoon_ccm}
\end{figure}

\section{Random networks with a constrained giant component}

The core idea of our approach is to build upon the classic configuration model but considering two types of connections: tagging some connections as critically important and therefore following specific rules, and others as less important, subcritical, connections.  This tagging procedure can be arbitrarily complex, but must result in a joint distribution $P(s,c)$ over the ``subcritical degree'' ($s$) and ``critical degree'' ($c$) of every node.

The critical constraint we explore involves identifying connections that point towards the giant component of a network.  Doing so will not only account for the size of that giant component, but also for some of its structure.  Consider two nodes of degree 100, both in the giant component, but one has 99 neighbors of degree 1, while the other has none and is part of multiple loops over the giant component.  Both nodes are connected to the giant component, but the second more robustly so, and we wish to account for this difference.

\subsection{Critical Configuration Model}
We assume an undirected network where every connection between nodes $i$ and $j$ can be thought of as two directed edges $(i,j)$ and $(j,i)$.  In this case, critical connections are tagged through the following procedure:
\begin{enumerate}
    \item Go over every directed edge $(i,j)$ in the network.
    \item Temporarily mask all edges pointing towards $i$, including $(j,i)$, to avoid backtracking.
    \item Ask whether the component downstream from $(i,j)$ is part of the largest connected component and greater than some threshold value (we use $1\%$ of system size).
    \item If yes, increase the critical degree $c_i$ of $i$ by 1. If not, increase the subcritical degree $s_i$ of $i$ by 1.
    \item Restore all directed edges pointing towards $i$.
    \item After going over all pairs, return a distribution $P(s,c)$.
\end{enumerate}

Based on the joint critical-subcritical degree distribution $P(s,c)$, one can then construct a random network coherent with the tagging procedure, such that critical degrees in the original network data remain critical in a randomized version of the data. We define this Critical Configuration Model as follows:
\begin{enumerate}
    \item Subcritical stubs connect to other stubs, but only towards nodes with excess critical degree 0.
    \item Critical stubs connect to other stubs, but only towards nodes with excess critical degree at least 1.
\end{enumerate}
Stubs are otherwise connected uniformly at random. Note that the second rule has a condition based on the excess critical degree of possible neighbors---i.e., the number of critical stubs that leave that node---which assures that there exists a giant component that contains all nodes with at least one critical degree, therefore preserving the size of the giant component.  The resulting model is illustrated in Fig.~\ref{fig:cartoon_ccm}.

\subsection{Size of the giant component}

We now consider the infinite random network ensemble defined by the aforementionned connection rules and the joint distribution $P(s,c)$ (with everything else random), and compute the expected size of the giant component using probability generating functions~\cite{newman2001structure}.

When selecting a node at random in the network, it will have subcritical degree $s$ and a critical degree $c$ drawn from a joint distribution $P(s,c)$, which is generated by
\begin{equation}
    G_0(x,y) = \sum_{s,c} P(s,c)x^sy^c \; .
\end{equation}
When following any type of stubs, the type of the stub through which we reach the neighboring node will be constrained by the excess degree of the reached node.  Regardless, the joint degree pair of the reached node will be biased by either its subcritical degree $s$ or critical degree $c$ as per the \textit{friendship paradox} since a random edge is 10 times more likely to reach a node of degree 10 than a node of degree 1.  To that effect, we introduce the neighbor generating function $f(x,y)$
\begin{equation}
    f(x,y) = \frac{\partial G_0(x,y)}{\partial x} + \frac{\partial G_0(x,y)}{\partial y} \; ,
\end{equation}
which is not normalized and therefore not a PGF.

More specifically, when following a subcritical stub, the connection rules defined previously state that we will reach a node proportionally to its number of subcritical stubs ($s$), or of critical stubs if it has only one ($c=1$).  The joint excess degree distribution of the reached node is therefore generated by $G_1^\mathrm{s}(x,y)$ and written as
\begin{equation}
    G_1^\mathrm{s}(x,y) = \dfrac{\sum_{s,c} P(s,c)\left(\delta_{c,1}x^s + s\delta_{c,0}x^{s-1}\right)}{\sum_{s,c} P(s,c)\left(\delta_{c,1} + s\delta_{c,0}\right)}
                        = \dfrac{f(x,0)}{f(1,0)} \; ,
\end{equation}
where $\delta_{a,b}$ is the Kronecker delta equal to 1 if $a$ equals $b$ and 0 otherwise.  Note that $G_1^\mathrm{s}(x,y)$ does not depend on $y$ since the critical excess degree of the reached node is 0, as per the connection rules.  Similarly, when following a critical degree, the connection rules state that we reach nodes proportionally to their total degree as long as they have excess critical degree of at least 1.  The joint excess degree distribution of these reached nodes is therefore generated by
\begin{align}
    G_1^\mathrm{c}(x,y) & = \dfrac{\sum_{s,c} P(s,c)\left(c\bar\delta_{c,1}x^sy^{c-1} + s\bar\delta_{c,0}x^{s-1}y^c\right)}{\sum_{s,c} P(s,c)\left(c\bar\delta_{c,1} + s\bar\delta_{c,0}\right)} \nonumber \\
                        & = \dfrac{f(x,y)-f(x,0)}{f(1,1)-f(1,0)} \;,
\end{align}
where $\bar\delta_{a,b} \equiv 1-\delta_{a,b}$.

\begin{figure}
    \centering
    \includegraphics[width=0.96\linewidth]{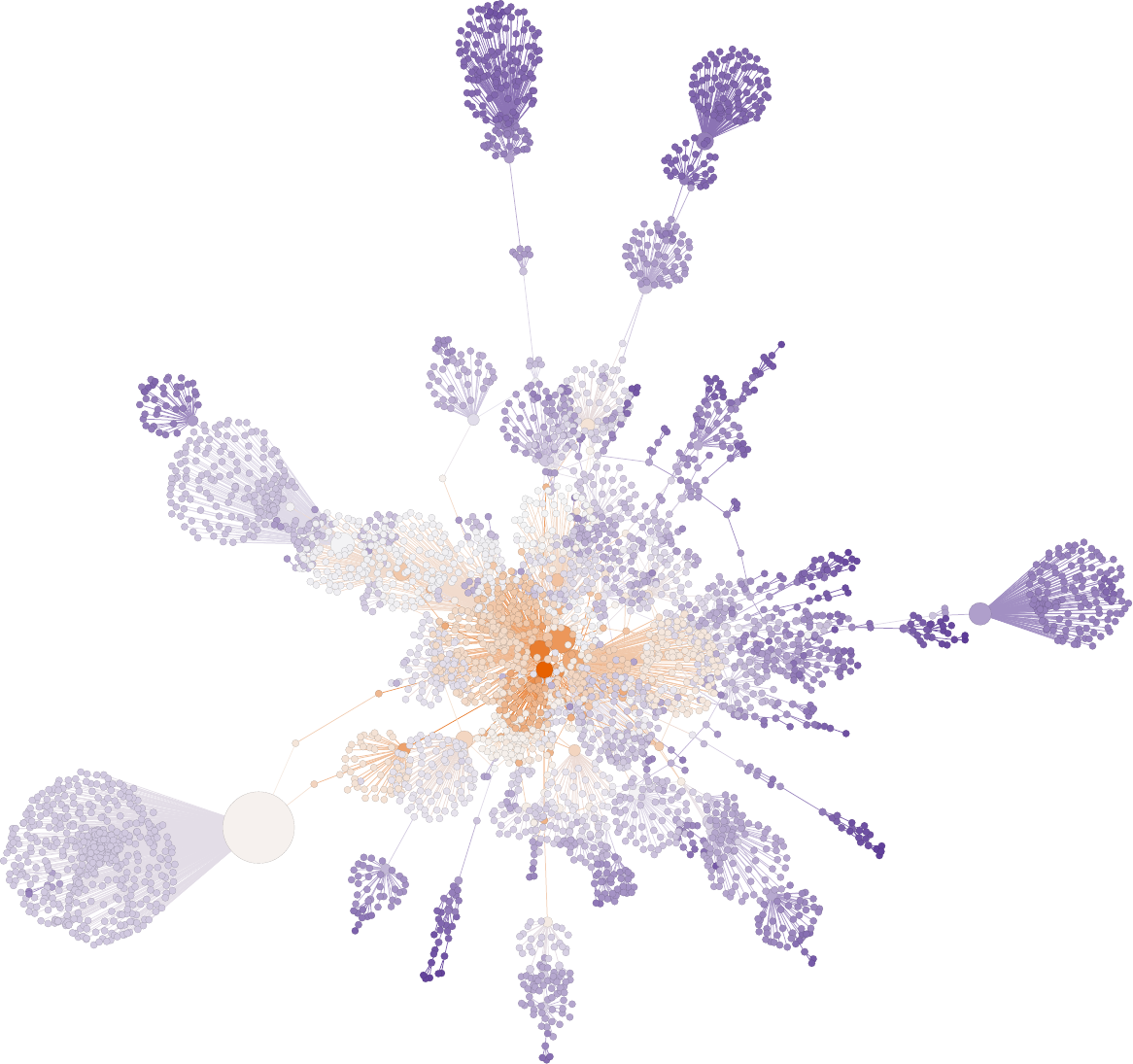}
    \includegraphics[width=0.96\linewidth]{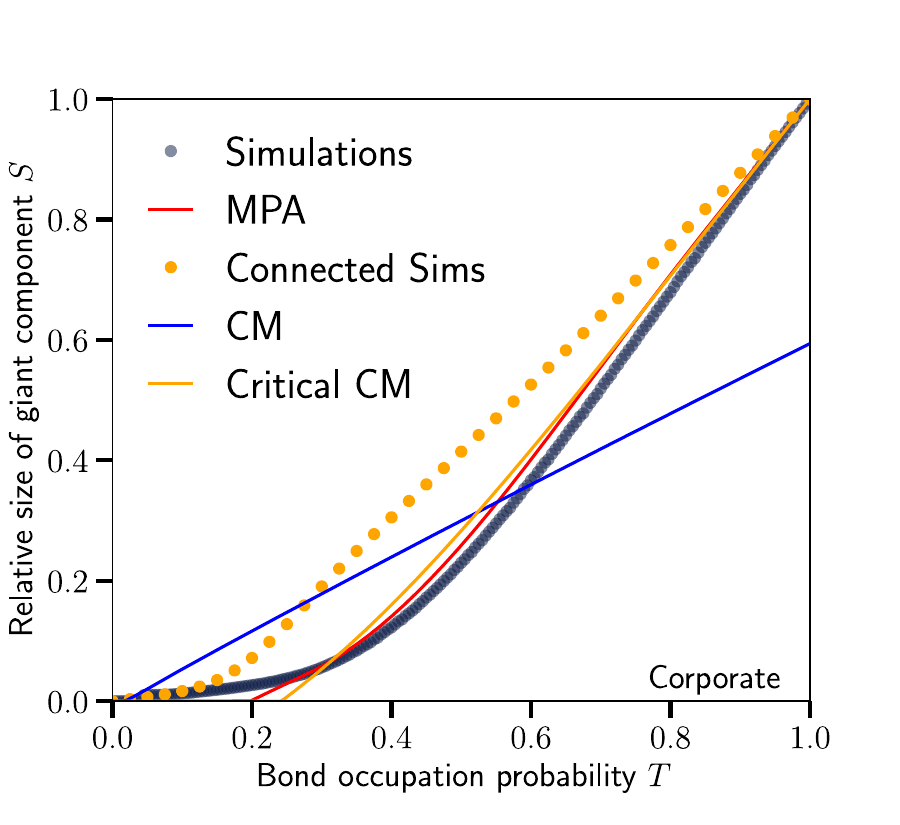}
    \caption{\textbf{(top)} The giant component of a corporate ownership network~\cite{norlen2002eva}.  Node size corresponds to degree (linearly from 1 to 552) and node color correspond to closeness centrality (inverse of the average distance to other nodes).  This is an interesting network structure in part because of its extremely skewed degree distribution, and because its most connected node has a much higher degree than its second most connected (552 vs 178) but the former is located in the periphery of the network while the latter is located in its core.  \textbf{(bottom)} We simulate percolation on this network and compare the results against 3 analytical models: the message passing approach (MPA; 9304 self-consistent equations), the configuration model (CM; 1 self-consistent equation) and our critical configuration model (Critical CM; 1 self-consistent equation).  We also compare the results against simulations on a connected subset of the configuration model (Connected Sims)~\cite{ring2020connected} to show that better models do more than capture the size of the connected component at $T=1$.}
    \label{fig:Corporate}
\end{figure}

With these equations in hand, we can then solve for the component size distribution of the network under a bond percolation process where every edge exists independently with probability $T$.  We denote $u_\mathrm{s}$ and $u_\mathrm{c}$ the probabilities that a subcritical or critical stub (respectively) does not lead to the giant component.  The first, $u_\mathrm{s}$, is fixed to 1 by construction whereas the second is the solution of the following self-consistent equation
\begin{equation} \label{eq:self_consistent_equation}
    u_\mathrm{c} = G_1^\mathrm{c}(1,1-T+Tu_\mathrm{c}) \; .
\end{equation}
In words, the probability that a critical stub does not lead to the giant component must be equal to the probability that the remaining stubs of the reached node do not lead to the giant component either.  This occurs if each critical stub \textit{leaving} the reached node corresponds to a removed edge (probability $1-T$) or, if the edge has not been removed, does not lead to the giant component either (probability $Tu_\mathrm{c}$).  References~\cite{newman2002spread, allard2015general, allard2019percolation} provide more details on this self-consistent argument.

Having solved Eq.~\eqref{eq:self_consistent_equation}, we may now compute the size of the giant component as the fraction of nodes for which at least one critical stub leads to it. Recalling that an individual critical stubs does not lead to the giant component with probability $1 - T + Tu_\mathrm{c}$, we find
\begin{align}
    S & = \sum_{s,c} P(s,c) \big[1 - (1 - T + Tu_\mathrm{c})^c\big] \nonumber \\
      & = 1 - G_0(1, 1 - T + Tu_\mathrm{c}) \; .
\end{align}
Notice that $u_\mathrm{c}=0$ is the only nontrivial solution of Eq.~\eqref{eq:self_consistent_equation} when $T=1$, meaning that the networks in our model will be connected with probability 1 in the limit of infinite networks in the absence of damage, as expected

\begin{figure*}
    \centering
    \includegraphics[width=0.47\linewidth]{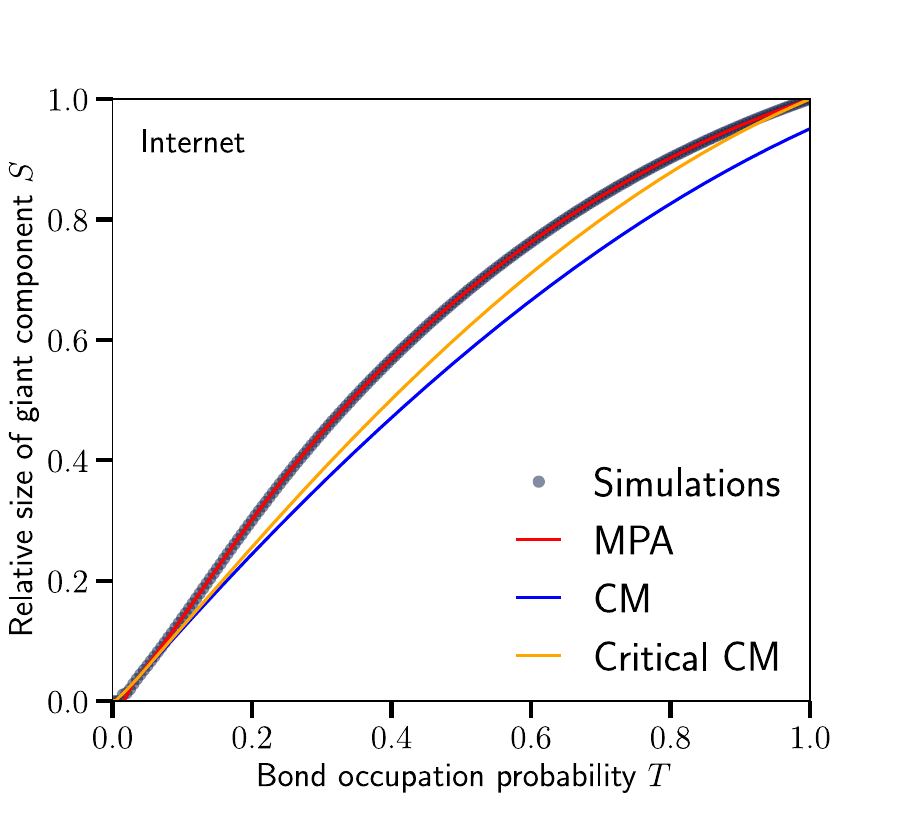}
    \includegraphics[width=0.47\linewidth]{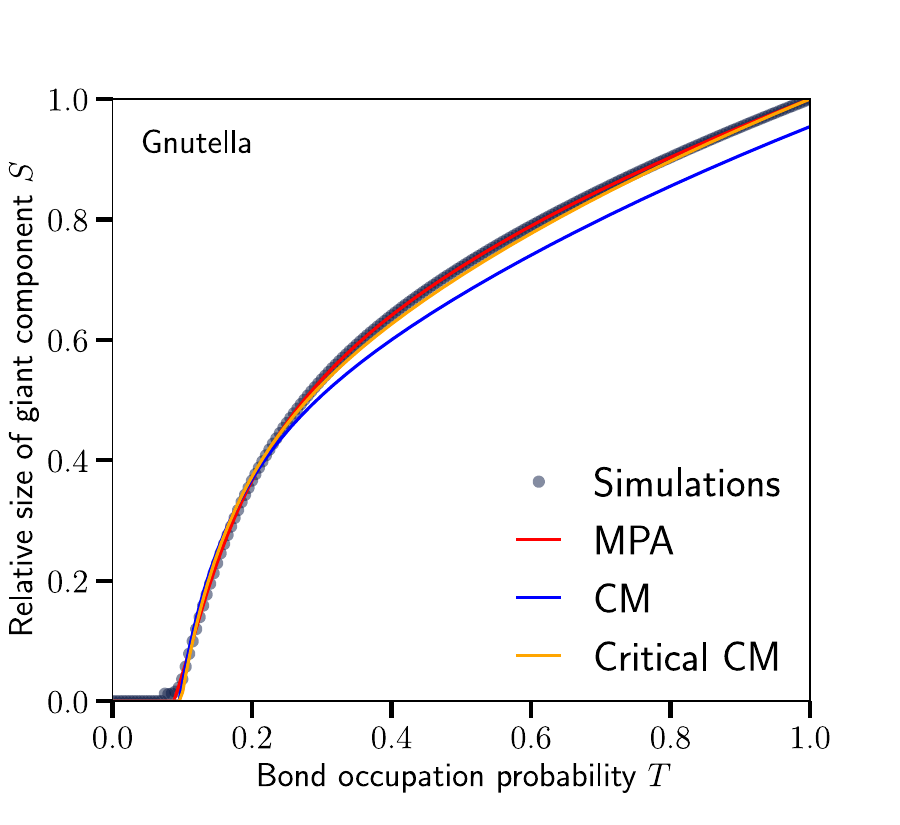}\\
    \includegraphics[width=0.47\linewidth]{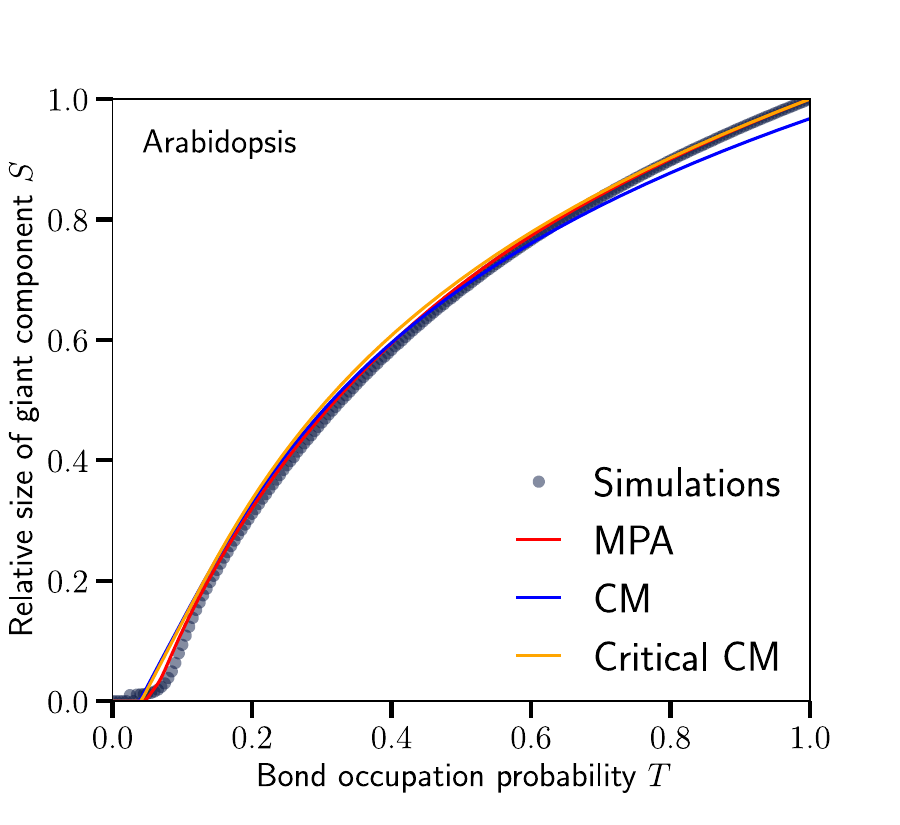}
    \includegraphics[width=0.47\linewidth]{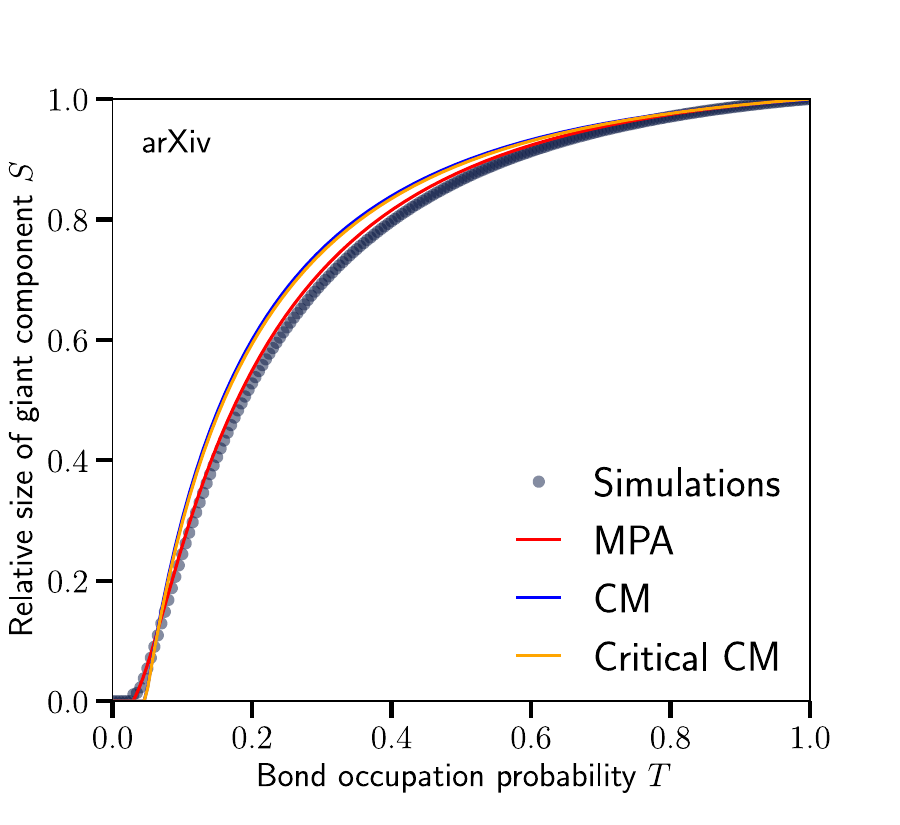}
    \caption{Percolation on the critical Configuration Model (Critical CM) and Configuration Model (CM) compared to simulations on the giant connected component of: \textbf{(top left)} the structure of the Internet at the level of autonomous systems~\cite{karrer2014percolation}, \textbf{(top right)} the peer-to-peer Gnutella network~\cite{matei2002mapping}, \textbf{(bottom left)} the Arabidopsis brain interactome~\cite{dreze2011evidence}, and \textbf{(bottom right)} a co-authorship network from the arXiv preprint archive~\cite{newman2001structure}.}
    \label{fig:multipanel}
\end{figure*}

\subsection{Connections with other models}

Most, if not all, random network models can be seen as mathematical frameworks used to compress complex network data based on some important features or constraints.  Our approach differ from common models in two ways.

First, instead of calibrating the model based on simple observations from network data (e.g., degree distribution) and validating the model by predicting the robustness of the corresponding network, the Critical CM combines the calibration and validation steps.  It uses information about how every node is embedded within the components found in the network data, and then attempts to extrapolate to perturbed version of the same network.

Second, calibrating the Critical CM requires more computational pre-processing than the CM.  It does not merely count edges around every node as in the CM, but tags them all based on some criterion for ``critical'' connections which can be costly to evaluate.

There is therefore a conceptual jump from the CM to the Critical CM since we now distinguish types of connections and now have to different types of stubs.  However, because the role of one type of stub is fixed by construction (recall that $u_\mathrm{s} = 1$), the resulting model is mathematically as complex as the CM as we only need to solve for a single polynomial self-consistent quantity.

By contrast, state-of-the-art approaches rely on a similar conceptual jump but greatly increase the complexity of the resulting mathematics.  The Message Passing Approach (MPA) is based on the idea that all stubs and all edges are distinguishable, and therefore tags the stub from node $i$ leading to node $j$ as a unique edge type $i\rightarrow j$~\cite{karrer2014percolation, allard2019accuracy}.  It then follows a similar calculation: Assuming an infinite number of nodes of any type (i.e. ignoring feedback through loops) whose degree sequence in types of stubs is explicitly given by the adjacency matrix of the true network, we can write a system of $2E$ self-consistent equations describing percolation on a network with $E$ edges.

The Critical CM therefore lies somewhere in-between the CM and MPA.  It can be seen as a fine-grained version of the configuration model where stubs are distinguished based on the structure of the non-backtracking component to which they lead.  Alternatively, the Critical CM can be seen as a coarse-graining of the MPA where we compress the $2E$ types of uniques connections into only two, again based on the non-backtracking component to which they lead.  By comparing the Critical CM to the CM and MPA, we hope to justify the development of more creative random network models.  In other words, we can, in theory, efficiently compress mathematical models by using network analysis as a pre-processing tools.

\section{Results}

We first compare the results of the Critical CM to other models and simulations based on the giant component of a common corporate ownership dataset, shown in Fig.~\ref{fig:Corporate}.  This network was chosen because the CM completely fails to match simulations of a percolation process, in large part because the giant component contains a gigantic hub in its periphery.  The CM completely overestimates the robustness of the giant component by underestimating the epidemic threshold, while conversely underestimating the final relative size of this giant component at $T=1$.  By comparing to simulations on a connected subset of the configuration model \cite{ring2020connected}, we can show that by capturing the complete size of the giant component, one also captures the higher than expected percolation threshold due to the prevalence of negative degree correlations required for this connectivity (hubs have to connect multiple low-degree nodes).  Interestingly, the Critical CM does a much better job at capturing the unique structure of this network by accounting not only for the size of the giant component but also for implicit correlations between the degree of nodes and their role in the cohesion of the giant component (the so-called ``critical degree'').  This highlights how our analytical model is not the same as a uniform connected subset of the CM, but an effective model that captures how the degree sequence of the network interacts with its giant component through the correlations between the degree of a node and its critical degree.  In fact, the Critical CM mimics results from the MPA very closely while decreasing the number of equations by a factor equal to twice the number of edges (1 versus 9304 self-consistent equations).

We further compare the Critical CM to the classic CM and to simulations for four other connected networks in Fig.~\ref{fig:multipanel}.  These networks were chosen to highlight how the previous results hold in technological network (e.g., the structure of the Internet) where the connectivity of the empirical networks might be surprising to the CM based on their degree distributions alone.  However, in denser networks such as brain connectomes and social networks, the Critical CM offers less of a gain and falls closer to the classic CM.  The usefulness of our approach therefore relies on the structure of interest, in this case the giant component, being surprising to classic random networks.

\section{Conclusion}

Random network models are used for multiple reasons across network science and beyond.  While many problems and dynamical systems are impossible to solve on a fixed network structure, they are often solvable on infinite random network ensembles.  When these models work well, we build intuition about what structural constraints matter the most for the problems and dynamics of interest.  This in turn also allows us to control for important features and therefore provide important null models for network analyses.  Indeed, we often need to ask how surprising is a network feature given some other metric.  That is why modularity or even degree assortativity control for degree distribution, using the CM as an underlying null model.  Through this pipeline, improvements in random network models improve much of network science.

Given that network scientists often care about global properties of network structures, such as their connectivity and robustness, we need random network models that can control for these features.  Unfortunately, most models are only built around simple local connection rules, leaving a significant gap in the network science toolbox.  While we certainly do not claim that the Critical CM introduced here should become a standard null model for network analyses and mathematical descriptions, it highlights the potential for new creative representation of network structures.  These can be models whose parametrization might be more complicated, involving computational pre-processing, but whose mathematical descriptions remain parsimonious.  In a nutshell, this line of research can be summarized as two simple questions: ``What features do we care about in a given network structure? And how can we best describe networks using these features?'' And when the answer to the first question is the global connectivity of a network, variations of the Critical CM presented here should prove useful.

\section*{Acknowledgments}

L.H.-D. acknowledges support from the Vermont Complex Systems Center and the College of Engineering and Mathematical Sciences at the University of Vermont. This work was supported in part by the Natural Sciences and Engineering Research Council of Canada (AA), the Sentinelle Nord program of Universit\'e Laval funded by the Canada First Research Excellence Fund (AA), and ERC grant No.~810115-DYNASET (M.P.).


\begin{thebibliography}{14}%
\makeatletter
\providecommand \@ifxundefined [1]{%
 \@ifx{#1\undefined}
}%
\providecommand \@ifnum [1]{%
 \ifnum #1\expandafter \@firstoftwo
 \else \expandafter \@secondoftwo
 \fi
}%
\providecommand \@ifx [1]{%
 \ifx #1\expandafter \@firstoftwo
 \else \expandafter \@secondoftwo
 \fi
}%
\providecommand \natexlab [1]{#1}%
\providecommand \enquote  [1]{``#1''}%
\providecommand \bibnamefont  [1]{#1}%
\providecommand \bibfnamefont [1]{#1}%
\providecommand \citenamefont [1]{#1}%
\providecommand \href@noop [0]{\@secondoftwo}%
\providecommand \href [0]{\begingroup \@sanitize@url \@href}%
\providecommand \@href[1]{\@@startlink{#1}\@@href}%
\providecommand \@@href[1]{\endgroup#1\@@endlink}%
\providecommand \@sanitize@url [0]{\catcode `\\12\catcode `\$12\catcode
  `\&12\catcode `\#12\catcode `\^12\catcode `\_12\catcode `\%12\relax}%
\providecommand \@@startlink[1]{}%
\providecommand \@@endlink[0]{}%
\providecommand \url  [0]{\begingroup\@sanitize@url \@url }%
\providecommand \@url [1]{\endgroup\@href {#1}{\urlprefix }}%
\providecommand \urlprefix  [0]{URL }%
\providecommand \Eprint [0]{\href }%
\providecommand \doibase [0]{http://dx.doi.org/}%
\providecommand \selectlanguage [0]{\@gobble}%
\providecommand \bibinfo  [0]{\@secondoftwo}%
\providecommand \bibfield  [0]{\@secondoftwo}%
\providecommand \translation [1]{[#1]}%
\providecommand \BibitemOpen [0]{}%
\providecommand \bibitemStop [0]{}%
\providecommand \bibitemNoStop [0]{.\EOS\space}%
\providecommand \EOS [0]{\spacefactor3000\relax}%
\providecommand \BibitemShut  [1]{\csname bibitem#1\endcsname}%
\let\auto@bib@innerbib\@empty
\bibitem [{\citenamefont {Fosdick}\ \emph {et~al.}(2018)\citenamefont
  {Fosdick}, \citenamefont {Larremore}, \citenamefont {Nishimura},\ and\
  \citenamefont {Ugander}}]{fosdick2018configuring}%
  \BibitemOpen
  \bibfield  {author} {\bibinfo {author} {\bibfnamefont {B.~K.}\ \bibnamefont
  {Fosdick}}, \bibinfo {author} {\bibfnamefont {D.~B.}\ \bibnamefont
  {Larremore}}, \bibinfo {author} {\bibfnamefont {J.}~\bibnamefont
  {Nishimura}}, \ and\ \bibinfo {author} {\bibfnamefont {J.}~\bibnamefont
  {Ugander}},\ }\bibfield  {title} {\enquote {\bibinfo {title} {Configuring
  {{Random Graph Models}} with {{Fixed Degree Sequences}}},}\ }\href {\doibase
  10.1137/16M1087175} {\bibfield  {journal} {\bibinfo  {journal} {SIAM Rev.}\
  }\textbf {\bibinfo {volume} {60}},\ \bibinfo {pages} {315--355} (\bibinfo
  {year} {2018})}\BibitemShut {NoStop}%
\bibitem [{\citenamefont {V{\'a}zquez}\ and\ \citenamefont
  {Moreno}(2003)}]{vazquez2003resilience}%
  \BibitemOpen
  \bibfield  {author} {\bibinfo {author} {\bibfnamefont {A.}~\bibnamefont
  {V{\'a}zquez}}\ and\ \bibinfo {author} {\bibfnamefont {Y.}~\bibnamefont
  {Moreno}},\ }\bibfield  {title} {\enquote {\bibinfo {title} {Resilience to
  damage of graphs with degree correlations},}\ }\href {\doibase
  10.1103/PhysRevE.67.015101} {\bibfield  {journal} {\bibinfo  {journal} {Phys.
  Rev. E}\ }\textbf {\bibinfo {volume} {67}},\ \bibinfo {pages} {015101}
  (\bibinfo {year} {2003})}\BibitemShut {NoStop}%
\bibitem [{\citenamefont {{H{\'e}bert-Dufresne}}\ \emph
  {et~al.}(2013)\citenamefont {{H{\'e}bert-Dufresne}}, \citenamefont {Allard},
  \citenamefont {Young},\ and\ \citenamefont
  {Dub{\'e}}}]{hebert2013percolation}%
  \BibitemOpen
  \bibfield  {author} {\bibinfo {author} {\bibfnamefont {L.}~\bibnamefont
  {{H{\'e}bert-Dufresne}}}, \bibinfo {author} {\bibfnamefont {A.}~\bibnamefont
  {Allard}}, \bibinfo {author} {\bibfnamefont {J.-G.}\ \bibnamefont {Young}}, \
  and\ \bibinfo {author} {\bibfnamefont {L.~J.}\ \bibnamefont {Dub{\'e}}},\
  }\bibfield  {title} {\enquote {\bibinfo {title} {Percolation on random
  networks with arbitrary k-core structure},}\ }\href {\doibase
  10.1103/PhysRevE.88.062820} {\bibfield  {journal} {\bibinfo  {journal} {Phys.
  Rev. E}\ }\textbf {\bibinfo {volume} {88}},\ \bibinfo {pages} {062820}
  (\bibinfo {year} {2013})}\BibitemShut {NoStop}%
\bibitem [{\citenamefont {Allard}\ and\ \citenamefont
  {{H{\'e}bert-Dufresne}}(2019{\natexlab{a}})}]{allard2019percolation}%
  \BibitemOpen
  \bibfield  {author} {\bibinfo {author} {\bibfnamefont {A.}~\bibnamefont
  {Allard}}\ and\ \bibinfo {author} {\bibfnamefont {L.}~\bibnamefont
  {{H{\'e}bert-Dufresne}}},\ }\bibfield  {title} {\enquote {\bibinfo {title}
  {Percolation and the {{Effective Structure}} of {{Complex Networks}}},}\
  }\href {\doibase 10.1103/PhysRevX.9.011023} {\bibfield  {journal} {\bibinfo
  {journal} {Phys. Rev. X}\ }\textbf {\bibinfo {volume} {9}},\ \bibinfo {pages}
  {011023} (\bibinfo {year} {2019}{\natexlab{a}})}\BibitemShut {NoStop}%
\bibitem [{\citenamefont {Albert}\ \emph {et~al.}(2000)\citenamefont {Albert},
  \citenamefont {Jeong},\ and\ \citenamefont {Barab{\'a}si}}]{albert2000error}%
  \BibitemOpen
  \bibfield  {author} {\bibinfo {author} {\bibfnamefont {R.}~\bibnamefont
  {Albert}}, \bibinfo {author} {\bibfnamefont {H.}~\bibnamefont {Jeong}}, \
  and\ \bibinfo {author} {\bibfnamefont {A.-L.}\ \bibnamefont {Barab{\'a}si}},\
  }\bibfield  {title} {\enquote {\bibinfo {title} {Error and attack tolerance
  of complex networks},}\ }\href {\doibase 10.1038/35019019} {\bibfield
  {journal} {\bibinfo  {journal} {Nature}\ }\textbf {\bibinfo {volume} {406}},\
  \bibinfo {pages} {378--382} (\bibinfo {year} {2000})}\BibitemShut {NoStop}%
\bibitem [{\citenamefont {Newman}(2002)}]{newman2002spread}%
  \BibitemOpen
  \bibfield  {author} {\bibinfo {author} {\bibfnamefont {M.~E.~J.}\
  \bibnamefont {Newman}},\ }\bibfield  {title} {\enquote {\bibinfo {title}
  {Spread of epidemic disease on networks},}\ }\href {\doibase
  10.1103/PhysRevE.66.016128} {\bibfield  {journal} {\bibinfo  {journal} {Phys.
  Rev. E}\ }\textbf {\bibinfo {volume} {66}},\ \bibinfo {pages} {016128}
  (\bibinfo {year} {2002})}\BibitemShut {NoStop}%
\bibitem [{\citenamefont {Newman}(2001)}]{newman2001structure}%
  \BibitemOpen
  \bibfield  {author} {\bibinfo {author} {\bibfnamefont {M.~E.~J.}\
  \bibnamefont {Newman}},\ }\bibfield  {title} {\enquote {\bibinfo {title} {The
  structure of scientific collaboration networks},}\ }\href {\doibase
  10.1073/pnas.021544898} {\bibfield  {journal} {\bibinfo  {journal} {Proc.
  Natl. Acad. Sci. U.S.A.}\ }\textbf {\bibinfo {volume} {98}},\ \bibinfo
  {pages} {404--409} (\bibinfo {year} {2001})}\BibitemShut {NoStop}%
\bibitem [{\citenamefont {Norlen}\ \emph {et~al.}(2002)\citenamefont {Norlen},
  \citenamefont {Lucas}, \citenamefont {Gebbie},\ and\ \citenamefont
  {Chuang}}]{norlen2002eva}%
  \BibitemOpen
  \bibfield  {author} {\bibinfo {author} {\bibfnamefont {K.}~\bibnamefont
  {Norlen}}, \bibinfo {author} {\bibfnamefont {G.}~\bibnamefont {Lucas}},
  \bibinfo {author} {\bibfnamefont {M.}~\bibnamefont {Gebbie}}, \ and\ \bibinfo
  {author} {\bibfnamefont {J.}~\bibnamefont {Chuang}},\ }\bibfield  {title}
  {\enquote {\bibinfo {title} {{{EVA}}: {{Extraction}}, {{Visualization}} and
  {{Analysis}} of the {{Telecommunications}} and {{Media Ownership
  Network}}},}\ }in\ \href@noop {} {\emph {\bibinfo {booktitle} {Proceedings of
  {{International Telecommunications Society}} 14th {{Biennial Conference}}
  ({{ITS2002}})}}}\ (\bibinfo {year} {2002})\ pp.\ \bibinfo {pages}
  {27---129}\BibitemShut {NoStop}%
\bibitem [{\citenamefont {Ring}\ \emph {et~al.}(2020)\citenamefont {Ring},
  \citenamefont {Young},\ and\ \citenamefont
  {{H{\'e}bert-Dufresne}}}]{ring2020connected}%
  \BibitemOpen
  \bibfield  {author} {\bibinfo {author} {\bibfnamefont {J.~H.}\ \bibnamefont
  {Ring}}, \bibinfo {author} {\bibfnamefont {J.-G.}\ \bibnamefont {Young}}, \
  and\ \bibinfo {author} {\bibfnamefont {L.}~\bibnamefont
  {{H{\'e}bert-Dufresne}}},\ }\bibfield  {title} {\enquote {\bibinfo {title}
  {Connected {{Graphs}} with a {{Given Degree Sequence}}: {{Efficient
  Sampling}}, {{Correlations}}, {{Community Detection}} and {{Robustness}}},}\
  }in\ \href {\doibase 10.1007/978-3-030-38965-9_3} {\emph {\bibinfo
  {booktitle} {Proceedings of {{NetSci-X}} 2020: {{Sixth International Winter
  School}} and {{Conference}} on {{Network Science}}}}}\ (\bibinfo {year}
  {2020})\ pp.\ \bibinfo {pages} {33--47}\BibitemShut {NoStop}%
\bibitem [{\citenamefont {Allard}\ \emph {et~al.}(2015)\citenamefont {Allard},
  \citenamefont {{H{\'e}bert-Dufresne}}, \citenamefont {Young},\ and\
  \citenamefont {Dub{\'e}}}]{allard2015general}%
  \BibitemOpen
  \bibfield  {author} {\bibinfo {author} {\bibfnamefont {A.}~\bibnamefont
  {Allard}}, \bibinfo {author} {\bibfnamefont {L.}~\bibnamefont
  {{H{\'e}bert-Dufresne}}}, \bibinfo {author} {\bibfnamefont {J.-G.}\
  \bibnamefont {Young}}, \ and\ \bibinfo {author} {\bibfnamefont {L.~J.}\
  \bibnamefont {Dub{\'e}}},\ }\bibfield  {title} {\enquote {\bibinfo {title}
  {General and exact approach to percolation on random graphs},}\ }\href
  {\doibase 10.1103/PhysRevE.92.062807} {\bibfield  {journal} {\bibinfo
  {journal} {Phys. Rev. E}\ }\textbf {\bibinfo {volume} {92}},\ \bibinfo
  {pages} {062807} (\bibinfo {year} {2015})}\BibitemShut {NoStop}%
\bibitem [{\citenamefont {Karrer}\ \emph {et~al.}(2014)\citenamefont {Karrer},
  \citenamefont {Newman},\ and\ \citenamefont
  {Zdeborov{\'a}}}]{karrer2014percolation}%
  \BibitemOpen
  \bibfield  {author} {\bibinfo {author} {\bibfnamefont {B.}~\bibnamefont
  {Karrer}}, \bibinfo {author} {\bibfnamefont {M.~E.~J.}\ \bibnamefont
  {Newman}}, \ and\ \bibinfo {author} {\bibfnamefont {L.}~\bibnamefont
  {Zdeborov{\'a}}},\ }\bibfield  {title} {\enquote {\bibinfo {title}
  {Percolation on {{Sparse Networks}}},}\ }\href {\doibase
  10.1103/PhysRevLett.113.208702} {\bibfield  {journal} {\bibinfo  {journal}
  {Phys. Rev. Lett.}\ }\textbf {\bibinfo {volume} {113}},\ \bibinfo {pages}
  {208702} (\bibinfo {year} {2014})}\BibitemShut {NoStop}%
\bibitem [{\citenamefont {Matei}\ \emph {et~al.}(2002)\citenamefont {Matei},
  \citenamefont {Iamnitchi},\ and\ \citenamefont {Foster}}]{matei2002mapping}%
  \BibitemOpen
  \bibfield  {author} {\bibinfo {author} {\bibfnamefont {R.}~\bibnamefont
  {Matei}}, \bibinfo {author} {\bibfnamefont {A.}~\bibnamefont {Iamnitchi}}, \
  and\ \bibinfo {author} {\bibfnamefont {P.}~\bibnamefont {Foster}},\
  }\bibfield  {title} {\enquote {\bibinfo {title} {Mapping the {{Gnutella}}
  network},}\ }\href {\doibase 10.1109/4236.978369} {\bibfield  {journal}
  {\bibinfo  {journal} {IEEE Internet Comput.}\ }\textbf {\bibinfo {volume}
  {6}},\ \bibinfo {pages} {50--57} (\bibinfo {year} {2002})}\BibitemShut
  {NoStop}%
\bibitem [{\citenamefont {Dreze}\ \emph {et~al.}(2011)\citenamefont {Dreze},
  \citenamefont {Carvunis}, \citenamefont {Charloteaux}, \citenamefont {Galli},
  \citenamefont {Pevzner}, \citenamefont {Tasan}, \citenamefont {Ahn},
  \citenamefont {Balumuri}, \citenamefont {Barabasi}, \citenamefont {Bautista},
  \citenamefont {Braun}, \citenamefont {Byrdsong}, \citenamefont {Chen},
  \citenamefont {Chesnut}, \citenamefont {Cusick}, \citenamefont {Dangl},
  \citenamefont {{de los Reyes}}, \citenamefont {Dricot}, \citenamefont
  {Duarte}, \citenamefont {Ecker}, \citenamefont {Fan}, \citenamefont {Gai},
  \citenamefont {Gebreab}, \citenamefont {Ghoshal}, \citenamefont {Gilles},
  \citenamefont {Gutierrez}, \citenamefont {Hao}, \citenamefont {Hill},
  \citenamefont {Kim}, \citenamefont {Kim}, \citenamefont {Lurin},
  \citenamefont {MacWilliams}, \citenamefont {Matrubutham}, \citenamefont
  {Milenkovic}, \citenamefont {Mirchandani}, \citenamefont {Monachello},
  \citenamefont {Moore}, \citenamefont {Mukhtar}, \citenamefont {Olivares},
  \citenamefont {Patnaik}, \citenamefont {Poulin}, \citenamefont {Przulj},
  \citenamefont {Quan}, \citenamefont {Rabello}, \citenamefont {Ramaswamy},
  \citenamefont {Reichert}, \citenamefont {Rietman}, \citenamefont {Rolland},
  \citenamefont {Romero}, \citenamefont {Roth}, \citenamefont {Santhanam},
  \citenamefont {Schmitz}, \citenamefont {Shinn}, \citenamefont {Spooner},
  \citenamefont {Stein}, \citenamefont {Swamilingiah}, \citenamefont {Tam},
  \citenamefont {Vandenhaute}, \citenamefont {Vidal}, \citenamefont {Waaijers},
  \citenamefont {Ware}, \citenamefont {Weiner}, \citenamefont {Wu},\ and\
  \citenamefont {Yazaki}}]{dreze2011evidence}%
  \BibitemOpen
  \bibfield  {author} {\bibinfo {author} {\bibfnamefont {M.}~\bibnamefont
  {Dreze}}, \bibinfo {author} {\bibfnamefont {A.-R.}\ \bibnamefont {Carvunis}},
  \bibinfo {author} {\bibfnamefont {B.}~\bibnamefont {Charloteaux}}, \bibinfo
  {author} {\bibfnamefont {M.}~\bibnamefont {Galli}}, \bibinfo {author}
  {\bibfnamefont {S.~J.}\ \bibnamefont {Pevzner}}, \bibinfo {author}
  {\bibfnamefont {M.}~\bibnamefont {Tasan}}, \bibinfo {author} {\bibfnamefont
  {Y.-Y.}\ \bibnamefont {Ahn}}, \bibinfo {author} {\bibfnamefont
  {P.}~\bibnamefont {Balumuri}}, \bibinfo {author} {\bibfnamefont {A.-L.}\
  \bibnamefont {Barabasi}}, \bibinfo {author} {\bibfnamefont {V.}~\bibnamefont
  {Bautista}}, \bibinfo {author} {\bibfnamefont {P.}~\bibnamefont {Braun}},
  \bibinfo {author} {\bibfnamefont {D.}~\bibnamefont {Byrdsong}}, \bibinfo
  {author} {\bibfnamefont {H.}~\bibnamefont {Chen}}, \bibinfo {author}
  {\bibfnamefont {J.~D.}\ \bibnamefont {Chesnut}}, \bibinfo {author}
  {\bibfnamefont {M.~E.}\ \bibnamefont {Cusick}}, \bibinfo {author}
  {\bibfnamefont {J.~L.}\ \bibnamefont {Dangl}}, \bibinfo {author}
  {\bibfnamefont {C.}~\bibnamefont {{de los Reyes}}}, \bibinfo {author}
  {\bibfnamefont {A.}~\bibnamefont {Dricot}}, \bibinfo {author} {\bibfnamefont
  {M.}~\bibnamefont {Duarte}}, \bibinfo {author} {\bibfnamefont {J.~R.}\
  \bibnamefont {Ecker}}, \bibinfo {author} {\bibfnamefont {C.}~\bibnamefont
  {Fan}}, \bibinfo {author} {\bibfnamefont {L.}~\bibnamefont {Gai}}, \bibinfo
  {author} {\bibfnamefont {F.}~\bibnamefont {Gebreab}}, \bibinfo {author}
  {\bibfnamefont {G.}~\bibnamefont {Ghoshal}}, \bibinfo {author} {\bibfnamefont
  {P.}~\bibnamefont {Gilles}}, \bibinfo {author} {\bibfnamefont {B.~J.}\
  \bibnamefont {Gutierrez}}, \bibinfo {author} {\bibfnamefont {T.}~\bibnamefont
  {Hao}}, \bibinfo {author} {\bibfnamefont {D.~E.}\ \bibnamefont {Hill}},
  \bibinfo {author} {\bibfnamefont {C.~J.}\ \bibnamefont {Kim}}, \bibinfo
  {author} {\bibfnamefont {R.~C.}\ \bibnamefont {Kim}}, \bibinfo {author}
  {\bibfnamefont {C.}~\bibnamefont {Lurin}}, \bibinfo {author} {\bibfnamefont
  {A.}~\bibnamefont {MacWilliams}}, \bibinfo {author} {\bibfnamefont
  {U.}~\bibnamefont {Matrubutham}}, \bibinfo {author} {\bibfnamefont
  {T.}~\bibnamefont {Milenkovic}}, \bibinfo {author} {\bibfnamefont
  {J.}~\bibnamefont {Mirchandani}}, \bibinfo {author} {\bibfnamefont
  {D.}~\bibnamefont {Monachello}}, \bibinfo {author} {\bibfnamefont
  {J.}~\bibnamefont {Moore}}, \bibinfo {author} {\bibfnamefont {M.~S.}\
  \bibnamefont {Mukhtar}}, \bibinfo {author} {\bibfnamefont {E.}~\bibnamefont
  {Olivares}}, \bibinfo {author} {\bibfnamefont {S.}~\bibnamefont {Patnaik}},
  \bibinfo {author} {\bibfnamefont {M.~M.}\ \bibnamefont {Poulin}}, \bibinfo
  {author} {\bibfnamefont {N.}~\bibnamefont {Przulj}}, \bibinfo {author}
  {\bibfnamefont {R.}~\bibnamefont {Quan}}, \bibinfo {author} {\bibfnamefont
  {S.}~\bibnamefont {Rabello}}, \bibinfo {author} {\bibfnamefont
  {G.}~\bibnamefont {Ramaswamy}}, \bibinfo {author} {\bibfnamefont
  {P.}~\bibnamefont {Reichert}}, \bibinfo {author} {\bibfnamefont {E.~A.}\
  \bibnamefont {Rietman}}, \bibinfo {author} {\bibfnamefont {T.}~\bibnamefont
  {Rolland}}, \bibinfo {author} {\bibfnamefont {V.}~\bibnamefont {Romero}},
  \bibinfo {author} {\bibfnamefont {F.~P.}\ \bibnamefont {Roth}}, \bibinfo
  {author} {\bibfnamefont {B.}~\bibnamefont {Santhanam}}, \bibinfo {author}
  {\bibfnamefont {R.~J.}\ \bibnamefont {Schmitz}}, \bibinfo {author}
  {\bibfnamefont {P.}~\bibnamefont {Shinn}}, \bibinfo {author} {\bibfnamefont
  {W.}~\bibnamefont {Spooner}}, \bibinfo {author} {\bibfnamefont
  {J.}~\bibnamefont {Stein}}, \bibinfo {author} {\bibfnamefont {G.~M.}\
  \bibnamefont {Swamilingiah}}, \bibinfo {author} {\bibfnamefont
  {S.}~\bibnamefont {Tam}}, \bibinfo {author} {\bibfnamefont {J.}~\bibnamefont
  {Vandenhaute}}, \bibinfo {author} {\bibfnamefont {M.}~\bibnamefont {Vidal}},
  \bibinfo {author} {\bibfnamefont {S.}~\bibnamefont {Waaijers}}, \bibinfo
  {author} {\bibfnamefont {D.}~\bibnamefont {Ware}}, \bibinfo {author}
  {\bibfnamefont {E.~M.}\ \bibnamefont {Weiner}}, \bibinfo {author}
  {\bibfnamefont {S.}~\bibnamefont {Wu}}, \ and\ \bibinfo {author}
  {\bibfnamefont {J.}~\bibnamefont {Yazaki}},\ }\bibfield  {title} {\enquote
  {\bibinfo {title} {Evidence for {{Network Evolution}} in an {{Arabidopsis
  Interactome Map}}},}\ }\href {\doibase 10.1126/science.1203877} {\bibfield
  {journal} {\bibinfo  {journal} {Science}\ }\textbf {\bibinfo {volume}
  {333}},\ \bibinfo {pages} {601--607} (\bibinfo {year} {2011})}\BibitemShut
  {NoStop}%
\bibitem [{\citenamefont {Allard}\ and\ \citenamefont
  {{H{\'e}bert-Dufresne}}(2019{\natexlab{b}})}]{allard2019accuracy}%
  \BibitemOpen
  \bibfield  {author} {\bibinfo {author} {\bibfnamefont {Antoine}\ \bibnamefont
  {Allard}}\ and\ \bibinfo {author} {\bibfnamefont {Laurent}\ \bibnamefont
  {{H{\'e}bert-Dufresne}}},\ }\href {\doibase 10.48550/arXiv.1906.10377} {\emph
  {\bibinfo {title} {On the accuracy of message-passing approaches to
  percolation in complex networks}}},\ \bibinfo {type} {Preprint}\ \bibinfo
  {number} {arXiv:1906.10377}\ (\bibinfo {year} {2019})\BibitemShut {NoStop}%
\end{thebibliography}
\end{document}